\documentclass[useAMS,usenatbib]{mn2e}
\usepackage{natbib}

\usepackage[dvips]{graphics}
\usepackage{epsfig}
\usepackage{amsmath}
\usepackage{aas_macros}

\def\mearth{{\rm\,M_\oplus}}

\def\msun{{\rm\,M_\odot}}

\def\micron{\mu {\rm m}}

\title[Vega's hot dust from planetesimal-driven migration]{Vega's hot dust from icy planetesimals scattered inward by an outward-migrating planetary system}

\author[Raymond \& Bonsor]{Sean N. Raymond$^{1,2,3}$\thanks{E-mail: rayray.sean@gmail.com} and Amy Bonsor$^4$\\
$^{1}$CNRS, Laboratoire d'Astrophysique de Bordeaux, UMR 5804, F-33270, Floirac, France \\
$^{2}$Univ. Bordeaux, Laboratoire d'Astrophysique de Bordeaux, UMR 5804, F-33270, Floirac, France.\\
$^{3}$NASA Astrobiology Institute's Virtual Planetary Laboratory.\\
$^{4}$School of Physics, University of Bristol H.H. Wills Physics Laboratory, Tyndall Avenue, Bristol, BS8 1TL, United Kingdom
} 

\begin{document}

\date{Accepted to MNRAS Letters}

\pagerange{\pageref{firstpage}--\pageref{lastpage}} \pubyear{2014}

\maketitle

\label{firstpage}

\begin{abstract}
Vega has been shown to host multiple dust populations, including both hot exo-zodiacal dust at sub-AU radii and a cold debris disk extending beyond 100~AU.  We use dynamical simulations to show how Vega's hot dust can be created by long-range gravitational scattering of planetesimals from its cold outer regions.  Planetesimals are scattered progressively inward by a system of 5-7 planets from 30-60 AU to very close-in.  In successful simulations the outermost planets are typically Neptune-mass.  The back-reaction of planetesimal scattering causes these planets to migrate outward and continually interact with fresh planetesimals, replenishing the source of scattered bodies.   The most favorable cases for producing Vega's exo-zodi have negative radial mass gradients, with sub-Saturn- to Jupiter-mass inner planets at 5-10 AU and outer planets of $2.5-20\mearth$.  The mechanism fails if a Jupiter-sized planet exists beyond $\sim15$~AU because the planet preferentially ejects planetesimals before they can reach the inner system.  Direct-imaging planet searches can therefore directly test this mechanism. 
\end{abstract}


\section{Introduction}

%, the primary photometric standard~\citep{hayes85},
Vega is the original rock star.  Observations dating back 30 years have shown a large excess in flux over the stellar photosphere at mid-infrared and longer wavelengths~\citep{aumann84,wilner02,sibthorpe10}.  Additional excesses have been measured in the $K$ band ($\sim 2 \micron$) of $\sim 1.3\%$ the stellar flux~\citep{absil06,defrere11} and at $10-30 \micron$ (excess of $\sim 7\%$)~\citep{su13}.  

The bright emission at $\lambda \ge 50 \micron$ comes from a resolved disk of more than 100~AU in radius~\citep{holland98,su05}.  Cold dust is thought to be generated by the collisional grinding of a planetesimal belt located between roughly 60 and 120~AU~\citep{su05,muller10}.  The excess at $2 \micron$ is attributed to hot ``exozodiacal'' dust at $\sim 0.2$~AU that is dominated by small grains and cannot be in collisional equilibrium because of the too-short timescales~\citep{defrere11}.  The warm excess at $10-30 \micron$ is spatially distinct from the hot emission and  may represent an asteroid belt at $\sim 14$~AU~\citep{su13}, although the flux is likely contaminated by the tail of the hot dust~\citep[e.g.][]{lebreton13} and, as we will explore below, perhaps by inward-scattered cold planetesimals.  
%Clumps have been detected in the dust distribution~\citep{wilner02}, a possible signpost of planets~\citep{wyatt06}.

Here we propose that Vega's hot exozodiacal dust may be supplied from its cold debris disk.  Planetesimals are transported inward by gravitational scattering by a system of planets.  The outermost planet scatters planetesimals inward and passes them off to the next planet until they reach the inner system; this is the same mechanism that delivers Jupiter-family comets to the inner Solar System~\citep{levison94} and is thought to be the main contributor to the Solar System's zodiacal dust~\citep{nesvorny10,rowanrobinson13}.  Planets on static orbits cannot sustain the inward-scattering rates needed to produce the observed exozodis because the outer planetesimal supply zones are depleted too rapidly~\citep{bonsor12}.  The observed exozodis can also not be caused by dynamical instabilities among systems of giant planets~\citep[see][]{raymond10} because the pulse of inward scattering is too short in duration~\citep{bonsor13}.  

For hot inner dust to come from the outer planetesimal belt, planets must continually interact with new material.  A solution is for the outermost planet to migrate outward.  For a given range of planet masses and disk properties, outward migration is naturally maintained by the gravitational back-reaction of inward-scattering of planetesimals~\citep{gomes04,kirsh09,ormel12}.  Planets that are too low-mass migrate rapidly but do not excite planetesimal eccentricities sufficiently to produce sufficient scattering.  Planetesimal driven migration naturally produces exozodis for a specific range of planet masses that depend on the disk's surface density~\cite[see][]{bonsor14}.  

Our simulations constitute a proof-of-concept experiment.  We determine the conditions needed for planetesimal-driven migration of a system of planets to generate a sufficient flux of inward-scattered planetesimals to roughly reproduce Vega's exozodi.  We consider the outer debris disk as the sole source of planetesimals.  We show that inward-scattered planetesimals can produce sufficient warm dust to explain the $10-30\micron$ excess without invoking steady-state collisional grinding in a dynamically cold asteroid belt.

\section{Simulations}

%\cite{muller10} showed that Vega's observed dust properties can be produced by an equilibrium collisional cascade within a belt of planetesimals.  Although several different disks provided adequate fits to the data, their best-fit disk extended from 62 to 120 AU and contained $46.7 \mearth$, having ground down by $\sim 20\%$ in mass over Vega's lifetime.  
%We performed simulations to show how planetesimal-driven migration in a planetary system can explain Vega's hot dust. In doing so we face many challenges regarding the uncertain conversion between the structure in large planetesimals and planets and the observed infra-red fluxes. 

Our goal was to produce the observed hot exozodiacal dust.  But the simulations also needed to match Vega's cold debris disk after 455 Myr of evolution. To simplify things, we focused on the best-fit model of \cite{muller10} that reproduces the observations with a disk extending from 62 to 120~AU containing $46.7 \mearth$, having ground down by $\sim 20\%$ in mass over Vega's lifetime.  The initial planetesimal disks were therefore somewhat more massive than the current one.  Our setup was also designed to trigger outward planetesimal-driven migration of the outermost planet, which plays a role in sculpting the inner edge of the debris disk.  We therefore started the planet significantly closer-in than the current planetesimal disk's inner edge.

Our fiducial setup had five planets placed between 5 and 30~AU, evenly spaced in orbital period ratio ($P_{i+1}/P_i \approx 1.95$).  Our choice of five planets produced systems that would indeed trigger the migration-scattering mechanism, although our setup was far from optimal~\citep[see][; more compact orbital configurations are more effective at scattering]{bonsor14}.  The number of planets is much less important than the fidelity of the chain of planets that passes planetesimals inward.  We varied the planets' masses between simulations as described below.  The planets' initial orbits had randomly-chosen eccentricities of less than 0.02 and randomly-chosen inclinations of less than $1^\circ$.  The planetesimal disk extended from 30 to 120 AU and contained $90 \mearth$. It followed an $r^{-1}$ surface density profile such that it started with $60 \mearth$ within the 60-120 AU belt of \cite{muller10}.  The disk contained 3000 planetesimals with randomly-chosen eccentricities of up to 0.01 and inclinations of up to $0.5^\circ$.  We adopted \cite{yoon10}'s stellar parameters for Vega (age of $455 \pm 13$~Myr and mass of $2.157 \pm 0.017 \msun$). 

To sustain an exozodi to its current age, the inner regions of the Vega system must be resupplied with material. In our simulations we tracked the flux of planetesimals scattered into the inner regions, defined as interior to 3~AU.  Any particle that entered within this radius was removed from the simulation. The details regarding the manner in which scattered planetesimals may evolve collisionally, or due to radiative forces and/or sublimation, potentially leading to a population of small dust grains similar to those thought to produce the exozodi emission, is a complex problem that we do not address here. Instead we consider a lower limit on the mass of material that must be delivered to the inner regions, and favor simulations that transport more material inwards. \cite{defrere11} estimated Vega's exozodi to contain $\sim 10^{-9} \mearth$ in dust with a characteristic lifetime of $\sim1$~yr.  We adopt $10^{-9} \mearth \, \mathrm{yr}^{-1}$ as a firm lower limit to the mass of material that must be delivered to the inner regions, and favor simulations that transport more material inwards.  On some figures a factor $f$ represents the fraction of inward-scattered planetesimals that must be converted to hot dust to match the observations (e.g., $f=10\%$ for a scattering rate of $10^{-8} \mearth \, \mathrm{yr}^{-1}$).

Each simulation was integrated for 500 Myr to 1 Gyr using the hybrid {\tt Mercury} integrator~\citep{chambers99}.  Planetesimals interacted gravitationally with the planets but not with each other.  Collisions were treated as inelastic mergers.  Particles were considered ejected and removed from the system if their radial distance exceeded $10^4$ AU.  The timestep was set to 200 days, designed to conserve energy to better than one part in $10^4$ outside 3 AU~\citep[see Appendix A in][]{raymond11}.  

\begin{figure}
  \begin{center} \leavevmode \epsfxsize=8.5cm\epsfbox{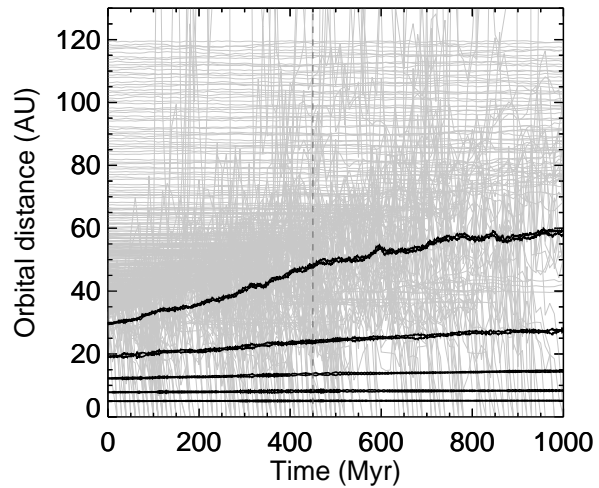}
%  \begin{center} \leavevmode \epsfxsize=8.5cm\epsfbox{at_all_100_50_25_10_10.eps}
    \caption[]{Evolution of the planets (periastron and apoastron in black) and planetesimals (semimajor axis in grey) in an example simulation. In radial order the planets' masses are 50, 25, 25, 10, and $10 \mearth$.  All planetesimals that entered within 3 AU are shown but only 1 of every 30 other planetesimals is shown.  Vega's estimated age is marked with a vertical dashed line.} 
     \label{fig:at_all}
    \end{center}
\end{figure}

Figure~\ref{fig:at_all} shows the evolution of a simulation with planets of 50, 25, 25, 10, and $10 \mearth$ (inner to outer).  The outermost planet migrated from 30 AU out to 59 AU.  The second- and third-outermost planets migrated from 19 to 28 AU and from 12 to 14.5 AU.  The inner two planets did not migrate substantially although they did cross the 2:1 mean motion resonance at $\sim300$~Myr, causing a modest increase in the planets' eccentricities.  Due to their large Hill radii, none of the planets accreted a substantial number of planetesimals.  The outermost planet was hit by the most planetesimals but accreted just $0.96 \mearth$. 

\begin{figure*}
  \begin{center} 
  \leavevmode 
    \epsfxsize=18cm\epsfbox{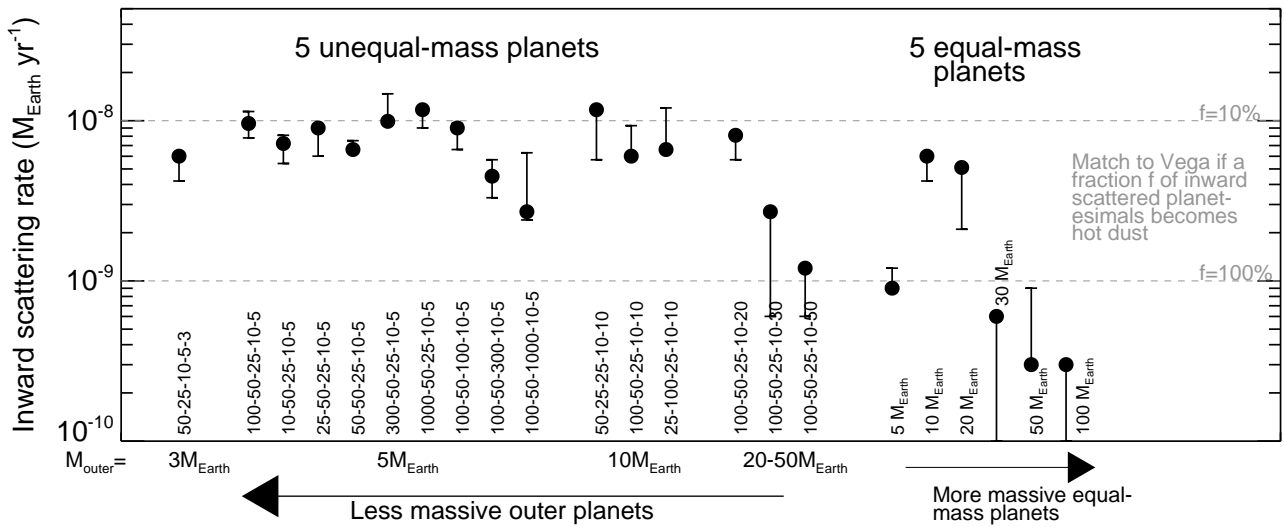}
\caption[]{Inward planetesimal-scattering rate of our simulations at Vega's age.  Symbols are 100 Myr averages from 400-500 Myr and error bars enclose the averages over the previous and subsequent 100 Myr intervals.  Simulations with unequal-mass planets are sorted by the mass of the outermost planet and then by the mass of the next-outermost planet and are labeled with the masses of all five planets (inner-to-outer).   The system from Fig.~\ref{fig:at_all} is ``50-25-25-10-10''.  Simulations with equal-mass planets are labeled by the individual planet mass.  The limits at $10^{-9} [10^{-8}] \mearth \,\mathrm{yr}^{-1}$ can produce Vega's observed hot dust if 100\% [10\%] of the mass in inward-scattered planetesimals turns to hot dust.}
    %The inward planetesimal-scattering rate for our simulations averaged over 100 Myr intervals.  The left panel shows simulations with five equal- mass planets in our fiducial disk.  The right panel shows simulations with five unequal-mass planets.  The unequal-mass simulations are grouped by the mass of the most massive planet in the system, usually the innermost one.  The vertical dashed line marks Vega's age.}
     \label{fig:rates}
    \end{center}
\end{figure*}
%The rate of planetesimals entering within 3 AU averaged over 100 Myr intervals.  The top left [top right] panel shows simulations with five equal- [unequal-] mass planets in our fiducial disk, the bottom left panel shows simulations in a lower-mass disk, and the bottom right shows simulations in a colder and lower-mass disk.  The unequal-mass simulations are grouped by the mass of the most massive planet in the system, usually the innermost one.  For the low-mass disk, simulations with equal-mass planets $M_p = 2.5-40 \mearth$.  The vertical dashed line marks Vega's age.

A total of $7.6 \mearth$ was scattered inside 3~AU in the simulation from Fig.\ref{fig:at_all} for an average of $7.6 \times 10^{-9}  \mearth \, {\mathrm yr}^{-1}$.  Almost all of the in-scattered planetesimals originated between the portion of the planetesimal disk initially between 30 and 60 AU.  The outer planet migrated through this portion of the disk and removed more than 80\% of the mass, mostly by scattering planetesimals to larger orbital radii (including ejecting them) or by scattering them inward toward the star.  A low-density belt of planetesimals survived at 35-50 AU.  These were scattered by the outermost planet but not strongly enough to encounter the next planet~\citep[see][]{raymond12}.  The disk is no longer as dynamically calm as the one from \cite{muller10}, yet it retains similar properties.  The median eccentricity of outer belt planetesimals is 0.05.  The amount of mass from 60-120 AU is $64 \mearth$ without accounting for collisional grinding, accelerated by the larger planetesimal eccentricities~\citep[e.g.][]{wyatt08}.   

Figure~\ref{fig:rates} shows the simulated inward-scattered flux at Vega's age.  A first conclusion is that systems with equal-mass planets can only produce Vega's exozodi for planet masses of $10-20 \mearth$.  Despite the low inward scattering rates for equal-mass planets, they are roughly an order of magnitude higher than corresponding simulations with planets on static orbits from \cite{bonsor12}.  Planetesimal-driven migration is clearly a key mechanism in maintaining a steady inward flux of planetesimals.  We note that planets evenly spaced in orbital period ratio are more efficient scatterers than planets separated by a fixed number of mutual Hill radii~\citep{bonsor14}.

Many simulations with unequal-mass planets had inward-scattering rates large enough to reproduce Vega's exozodi (middle panel).  The successful systems are broadly characterized by negative radial mass gradients.  The inner planet is $50 \mearth$ to (super-) Jupiter-mass, the outer 1-2 planets are $5-20 \mearth$, and the planets in between are $10-50 \mearth$.  The exact ordering of planets is not important as long as this general trend is followed. The simulation from Fig.~\ref{fig:at_all} has one of the highest inward-scattering rates, of $1.2 \times 10^{-8} \mearth \,\mathrm{yr}^{-1}$.  The unequal-mass systems that did not generate large inward-scattering rates suffered from one of two problems.  Either the mass of their outermost planet was not between $5 \mearth$ and $20 \mearth$, or the system contained a gas giant 

\subsection{Limits on the presence of a Jupiter-mass planet}
\begin{figure}
  \begin{center} \leavevmode \epsfxsize=8.5cm\epsfbox{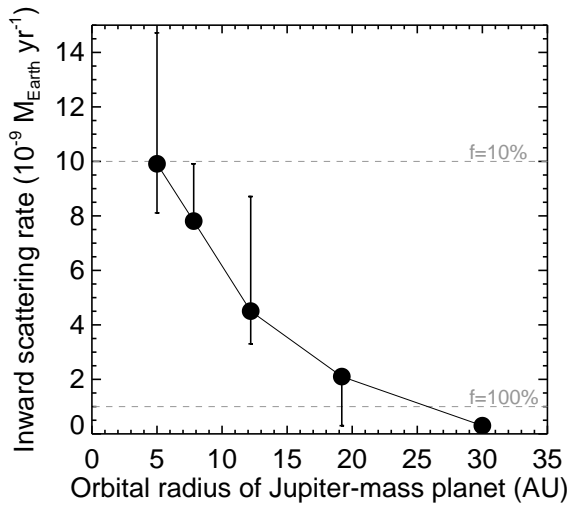}
%  \begin{center} \leavevmode \epsfxsize=8.5cm\epsfbox{vega_rates_mgrad_jup_color.eps}
%  \begin{center} \leavevmode \epsfxsize=8.5cm\epsfbox{vega_rates_mgrad_jup.eps}
    \caption[]{Inward scattering rate at Vega's age for systems in which each of the five planets in a fiducial simulation was successively replaced with a Jupiter-mass planet.  The symbols are averages from 400-500 Myr and the error bars encompass rates over the two previous and successive 100 Myr intervals.  The horizontal lines mark the rates for which 10\% and 100\% of in-scattered planetesimals must be converted to hot dust to match the observations.} 
     \label{fig:jup}
    \end{center}
\end{figure}

We tested the effect of a Jupiter-mass planet on exozodi production.  We started from a fiducial simulation with planet masses of 100, 50, 25, 10, and $5 \mearth$ (inner to outer).  The simulation had an inward scattering rate of $9.6 \times 10^{-9} \mearth \,\mathrm{yr}^{-1}$.  We performed five additional simulations. In each simulation we increased the mass of one of the five planets to one Jupiter-mass.  Figure~\ref{fig:jup} shows the inward scattering rate in these systems.  Increasing the mass of the innermost or second-innermost planet did not have a large effect on the scattering rate.  For a more distant Jupiter the rate of inward-scattered planetesimals was significantly depressed.  The rate was cut in half with a Jupiter in the middle of the system (at 12.2 AU).  The rate drops is borderline for a Jupiter at 19.2 AU and below the critical limit at 30 AU.  

With its high escape speed ($60 \, \mathrm{km \, s^{-1}}$), a Jupiter-mass planet ejects planetesimals more efficiently than it passes them inward.  It disrupts the supply chain between the outer and inner planetary systems.  The mechanism of inward scattering is thwarted by a gas giant planet beyond $\sim15$~AU.  This is testable: no gas giant planet can exist beyond $\sim 15$~AU for our mechanism to operate.  Current observations constrain the mass of a planet to be less than $3-10 {\mathrm M}_J$ beyond $\sim 10$~AU~\citep{itoh06,marois06,heinze08}.  New direct-imaging instruments are coming online soon, specifically {\it SPHERE}~\citep{beuzit08} and {\it Gemini Planet Imager}~\citep{macintosh08}.  Observations may directly constrain our model in the coming years.  

\subsection{Testing different planetesimal belt configurations}
To show that the mechanism is robust we ran two additional sets of simulations with different planetesimal disk properties.  The first set assumed a lower-mass disk with $50 \mearth$ initially contained between 30 and 120 AU.  The second set assumed a lower-mass and colder disk containing $50 \mearth$ between 45 and 120 AU.  Each disk contained 3000 planetesimals and the same orbital distribution as in the fiducial disk.  For the lower-mass disk we ran simulations with planets with similar mass gradients to the most successful simulations in the fiducial disk as well as equal-mass planets with $M_p = 2.5 - 40 \mearth$.  For the cold disk we kept the innermost planet at 5 AU while maintaining dynamical contact with the outer disk.  We placed the planets between 5 and 45 AU, again evenly spaced in orbital period ratio.  In some simulations we included 5 planets with $P_{i+1}/P_i = 2.3$ and in others 7 planets with $P_{i+1}/P_i = 1.7$.  

The simulations in cold disks produced inward scattering rates as high as $7 \times 10^{-9} \, \mearth \,{\mathrm yr^{-1}}$.  The most successful simulations had 7 planets including an Jupiter-Saturn like inner pair and outer planets with masses between $2.5 \mearth$ and $10 \mearth$.  Although 5-planet systems generally had much lower inward scattering rates, the most successful 5 planet-system -- with planet masses of 300, 100, 50, 25 and $10 \mearth$ from innermost to outermost -- had a comparable scattering rate to the successful 7-planet systems.  In the low-mass disk only two simulations had high enough inward scattering rates to match the $10^{-9} \, \mearth \,{\mathrm yr^{-1}}$ needed  to generate Vega's exozodi.  Both successful simulations had negative radial mass gradients, with (inner to outer) planets masses of 300, 100, 50, 25, $10 \mearth$, and 100, 50, 25, 10, $5\mearth$.  

\subsection{Matching observations}
%Simulations with large inward-scattering rates do not retain a smooth planetesimal distribution.  
We now estimate the effect that scattered planetesimals have on the observations.  Transforming a planetesimal population into a dust distribution is a complex process~\citep[e.g.][]{dominik03,wyatt08,krivov10}.  We use a simplified approach.  Our goal is not to perfectly reproduce the observations but to verify that observations of our simulated planetary systems would not differ enormously from those of Vega.  We used the code from \cite{raymond11,raymond12}, which follows the procedure of \cite{booth09}, based on the model of \cite{wyatt07b}.  

Each simulated planetesimal was assumed to represent a distribution of objects following a collisional size-frequency with $dN/dD \propto D^{-3.5}$~\citep{dohnanyi69}.  Each planetesimal is heated by the star and cools as a blackbody.  We took into account Vega's stellar properties~\citep{yoon10} to derive an approximate blowout size of $D_{bl} \approx 0.8 \left(L_\star/L_\odot \right) \left(M_\star/M_\odot \right)^{-1} \approx 15 \micron$. The surface area per unit mass of our collisional distribution is $\sigma/M = 0.07 \, {\mathrm AU^2} \, \mearth^{-1}$ (assuming $D_{max} = 2000$~km).  The collisional timescale $t_{coll}$ of the largest bodies was calculated and the planetesimal mass was decreased in time as $M_p(t) \propto \frac{1}{1+t/t_{coll}}$.  This is not strictly consistent with the dynamical integration in which the planetesimal mass did not change, but the cumulative decrease in $M_p$ is relatively small (typically 10-20\%).  We assumed a fixed value of $Q_D^\star = 200 \, {\mathrm J \, kg^{-1}}$~\citep{wyatt07b}.  This does not take into account realistic dust emission, so fluxes are to be trusted within an order of magnitude~\citep{booth09}.  
%We therefore do not expect a perfect fit with the data.

\begin{figure}
  \begin{center} \leavevmode \epsfxsize=8.5cm\epsfbox{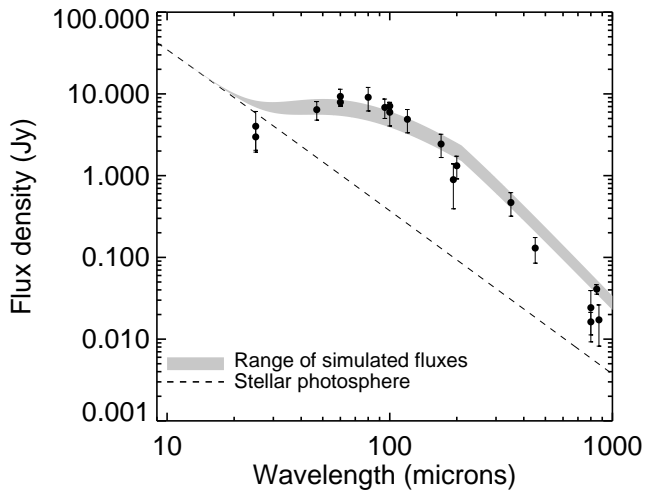}
%  \begin{center} \leavevmode \epsfxsize=8.5cm\epsfbox{SED_mgrad_cold_lowmass.eps}
%  \begin{center} \leavevmode \epsfxsize=8.5cm\epsfbox{SED_mgrad_cold.eps}
    \caption[]{Vega's observed spectral energy distribution (SED)~\citep[data compiled from Table 1 in][]{muller10} compared with SEDs generated by our simulations. The shaded region encloses the fluxes generated by three simulations that produced large inward-scattering fluxes: one in the fiducial disk (the upper bound), one in the cold disk and one in the low-mass disk. } 
     \label{fig:SED}
    \end{center}
\end{figure}

Figure~\ref{fig:SED} shows that the spectral energy distribution (SED) of our simulations provides a decent match to Vega's dust flux in the mid-infrared.  There is a small excess in the fiducial simulations (the upper bound of the shaded area in Fig.~\ref{fig:SED}) at $25 \micron$ and $47 \micron$.  The excess comes from planetesimals scattered inward on Centaur-like orbits.  These planetesimals are hotter than those in the outer disk and contaminate the SED starting at slightly shorter wavelength than the observed infrared excess.  They also generate warm dust at a similar level to the observations.  Inward-scattered planetesimals contribute a few to 10\% of the stellar flux at $10-30 \micron$, the same order of magnitude as \cite{su13}'s claimed asteroid belt.  In-scattered planetesimals may account for both Vega's warm and hot dust populations.  However, we caution that the true dust production of these planetesimals is highly uncertain.  As they may not be in collisional equilibrium as assumed, inward-scattered planetesimals may produce less dust than in our simple calculations. In addition, any terrestrial planets inside 3~AU would affect the long-term dynamics of both inward-scattered planetesimals and in-spiraling dust.

The dust SEDs from the successful simulations in cold- and low-mass disks somewhat underestimated the flux at $50-100 \micron$ because the planetesimal mass was somewhat below the expected one (the lower bound of the shaded region in Fig.~\ref{fig:SED}).  

%A careful consideration of the SED observations could further constrain Vega's planetary system, but that analysis is beyond the scope of this paper.

%Figure~\ref{fig:SED} shows that the spectral energy distribution (SED) of our fiducial simulation provides a decent match to Vega's observed dust flux throughout the mid-infrared.  There is a small excess in the simulation at $25 \micron$ and $47 \micron$.  The outer disk in the fiducial simulation is a reasonable representation of \cite{muller10}'s 62-120~AU equilibrium collisional disk.  The excess emission the simulation comes from planetesimals that are closer-in, many of which are in the process of being scattered inward on Centaur-like orbits.  These planetesimals are bright because they are hotter than those in the outer disk, and they contaminate the SED starting at slightly shorter wavelength than the observed infrared excess.    %We suspect that a more careful consideration of the SED observations could further constrain Vega's planetary system, but that analysis is beyond the scope of this paper.

\section{Discussion}

We have shown that Vega's exozodi can be explained by a system of planets with orbital radii of roughly 5-60 AU.  Planets scatter planetesimals from the outer planetesimal disk into the inner planetary system.  The outermost planets undergo plantesimal-driven migration outward into the disk, continually encounter fresh material and can sustain a substantial inward-scattering rate for the age of the system and beyond~\citep[see][for details of this process]{bonsor14}.  
%This mechanism is robust within the different tested planetesimal disks. 

For the mechanism to operate the system must meet the following requirements.  First, the outermost 1-2 planets must have masses such that planetesimal-driven migration is triggered and sustained.  Successful simulations had $M_{outer} = 2.5-20 \mearth$.  The inner planets' masses have only a minor effect, although the highest scattering rates were for simulations with higher-mass (Saturn- to Jupiter-mass) planets closer-in.  Second, while the inner system may host a gas giant, a $\sim$Jupiter-mass planet cannot orbit beyond $\sim15$~AU as it too-efficiently ejects planetesimals from the system and breaks the inward-scattering chain (see Fig.~\ref{fig:jup}).  This prediction does not violate current constraints~\citep{itoh06,marois06,heinze08} and may be tested with upcoming instruments.  Third, our model requires that a substantial fraction of the mass in planetesimals scattered within 3~AU must end up as fine-grained dust.  This fraction is $\sim$10-30\% for the simulations we have deemed successful.  If we take our simplified dust flux calculation at face value, our simulations provide a modestly good match to the observed SED (see Fig.~\ref{fig:SED}).  Although our simulations somewhat overestimate the flux at $25-50 \micron$, dust produced by in-scattered planetesimals may offer an alternate explanation for the $10-30\micron$ excess detected by \cite{su13}.  Planetesimal-driven migration of a system of 5-7 planets in a lower-mass disk and colder disk can also yield reasonable inward-scattering rates and SEDs.

\vskip .1in
\noindent We thank M. Booth and J.-C. Augereau for interesting discussions.  A.B. acknowledges the support of the ANR-2010 BLAN-0505- 01 (EXOZODI) and funding from NERC.  This work was performed as part of the NASA Astrobiology Institute's Virtual Planetary Laboratory Lead Team, supported by the NASA under Cooperative Agreement No.  NNA13AA93A.  This paper was entirely written while S.N.R. was standing and is therefore a ``sit-free'' paper.

%\bibliographystyle{mn2e_b}
%\bibliography{refs.bib}

\end{document}